\documentclass[prl,twocolumn,showpacs]{revtex4}

\usepackage{graphicx}

\usepackage{amsmath,amssymb}
\usepackage{bm}
\usepackage{cancel}

\newcommand{\calA}{\mathcal{A}}

\newcommand{\calP}{\mathcal{P}}
\newcommand{\calL}{\mathcal{L}}

\newcommand{\1}{\eta^{(1)}}
\newcommand{\2}{\eta^{(2)}}
\newcommand{\3}{\eta^{(3)}}
\newcommand{\auno}{a^{(1)}}
\newcommand{\atres}{a^{(3)}}

\newcommand{\eq}{\text{eq}}

\begin{document}
\title{Sawtooth patterns in biomolecules force-extension curves: an equilibrium-statistical-mechanics theory}
\author{A.\ Prados$^1$, A.\ Carpio$^2$, and L.\ L.\ Bonilla$^3$}
\affiliation{$^1$ F\'{\i}sica Te\'{o}rica, Universidad de Sevilla,
Apartado de Correos 1065, E-41080, Sevilla, Spain}
\affiliation{$^2$ Departamento de Matem\'atica Aplicada, Universidad Complutense de Madrid, 28040 Madrid, Spain}
\affiliation {$^3$G. Mill\'an Institute, Fluid Dynamics, Nanoscience and Industrial
Mathematics, Universidad Carlos III de Madrid, 28911 Legan\'es, Spain}

\date{\today}
\begin{abstract}
    We analyze the force-extension curve for a general class of systems, which are described at the mesoscopic level by a free energy depending on the extension of its components. Similarly to what is done in real experiments, the total length of the system is the controlled parameter. This imposes a global constraint in the minimization procedure leading to the equilibrium values of the extensions. As a consequence, the force-extension curve has multiple branches in a certain range of forces. The stability of these branches is governed by the free energy: there are a series of first-order phase transitions at certain values of the total length, in which the free energy itself is continuous but its first derivative, the force, has a finite jump. This behavior is completely similar to the one observed in real experiments with biomolecules like proteins, and other complex systems.

\end{abstract}
\maketitle

Nowadays technological advances allow manipulation of single molecules with sufficient precision to study many mechanical, kinetic and thermodynamic properties thereof. Recent reviews of techniques used in single-molecule experiments (SMEs), such as laser optical tweezers or atomic force microscopy (AFM), can be found in \cite{rit06jpcm,HyD12,MyD12}. For instance, in AFM experiments, a single molecule is stretched at a constant rate between the tip of a microscopic cantilever and a flat, gold-covered substrate, whose position can be externally controlled \cite{HyD12,MyD12,FOCMyF99,COFMBCyF99,FMyF00}. A typical outcome is a force-extension ($F-L$) curve characterizing the molecule elasticity. In principle, this $F-L$ curve is obtained in an out-of-equilibrium situation. Nevertheless, it has been used to recover free energy differences and to reconstruct the free energy landscape \cite{GRyB03,CRJSTyB05,JMMyR09,GVNYWyW11,LDSTyB02}, by means of the so-called transient fluctuation theorems derived by Jarzynski \cite{Ja97}, Crooks \cite{Cr00}, and extended by Hummer and Szabo \cite{HyS01}.

SMEs with proteins \cite{FOCMyF99,COFMBCyF99,FMyF00,HyD12,MyD12} or nucleic acids \cite{rit06jpcm,LDSTyB02,GRyB03,CRJSTyB05,JMMyR09,GVNYWyW11,LOSTyB01} have some key fingerprints in their $F-L$ curves. When a molecule is pulled, there is a critical value of the force (usually in the pN regime) at which it unfolds. Moreover,  there is a sudden decrease in the force that accompanies the unfolding, the so-called force-rip. The unfolding process often comprises several steps, because the molecule consists of several units that unfold one by one. Then a typical sawtooth pattern appears in the $F-L$ curve. This pattern is particularly clean for modular proteins \cite{HyD12,MyD12,FOCMyF99,COFMBCyF99,FMyF00}, although it has also been observed in nucleic acids \cite{rit06jpcm,HBFSByR10}.
Modular proteins are polyproteins comprising several (typically $6$ to $12$) repeats of identical protein folds, joined together by short peptide linkers. An example is the polyprotein I$27_8$, which is composed of eight copies of immunoglubulin domain $27$ from human cardiac titin \cite{FMyF00}. Identical domains have the same mechanical properties, which simplify the analysis of the polyprotein elastic response.

To be specific, we will focus on modular proteins throughout our work, but most of the ideas in our analysis can be extended (with the appropriate change of terminology) to other biomolecules as nucleic acids \cite{rit06jpcm,LDSTyB02,GRyB03,CRJSTyB05,JMMyR09,GVNYWyW11}. Similar phenomena occur in quite different physical systems such as lithium-based devices \cite{dre10,dre11}, storage of air in interconnected systems of rubber balloons \cite{dre11cmt}, or weakly coupled semiconductor superlattices \cite{BGr05,BT10}. In all these cases, the system can be thought to be composed of a number of similar bistable units, whose individual states are determined by a long-range interaction introduced by a global constraint. In biomolecules, this constraint is the imposed value of the system length, which is usually the control parameter.

In this Letter, we put forward a general class of models that are described by a Landau-like free energy \cite{landau} at the mesoscopic level. This free energy depends on the extension of the units composing the system at hand. For the sake of simplicity, we assume that the interactions between units is negligible. Then the free energy of the system is given by the sum of the free energies of its components. This free energy gives the equilibrium probability of finding the system in a certain configuration of its units extensions. We investigate the equilibrium $F-L$ curve under length controlled conditions. As a consequence, the force extension curve has multiple branches in a region of metastability thereby giving rise a sawtooth pattern with force rips completely similar to the experimentally observed one. The minimum value  of the force at the rips increases with the number of unfolded domains, even if the latter are completely identical.

Let us consider a modular protein that comprises $N$ domains, described at the mesoscopic level by their extensions $\eta_j$, $j=1,\ldots,N$. When isolated, the free energy $a_j$ of the $j$-th module depends on $\eta_j$ and also on a certain set $Y$ of intensive parameters (temperature $T$, pressure $p$, etc.\/) We consider the simple polynomial form $a_j(\eta_j;Y)=F_{c,j}(Y) \eta_j - \alpha_j(Y) \eta_j^2+\beta_j(Y) \eta_j^4,$ for its free energy, where $F_{c,j}$, $\alpha_j$ and $\beta_j$ are positive functions of $Y$. This  Landau-like expression suffices for the analysis we carry out here.
Moreover, we restrict ourselves to the \textit{homogeneous} situation in which all the modules are identical, $F_{c,j}=F_c$, $\alpha_j=\alpha$, $\beta_j=\beta$, $j=1,\ldots,N$, Thus $a_j=a$, independent of $j$, with
\begin{equation}\label{1}
  a(\eta_j;Y)=F_{c}(Y) \eta_j - \alpha(Y) \eta_j^2+\beta(Y) \eta_j^4.
\end{equation}
We assume that the interaction between modules can be neglected, so that the total free energy of the system can be written as
\begin{subequations}\label{2}
\begin{equation}\label{2a}
  A(\bm{\eta};Y)=\sum_{j=1}^N a(\eta_j;Y), \quad \bm{\eta}=\{\eta_1,\ldots,\eta_N\}.
\end{equation}
In order to mimic length controlled real experiments, we impose $\sum_{j=1}^N \eta_j=L$, where $L$ is the system length measured with respect to some reference value. The probability of finding the system in a certain extension configuration $\eta$ is given by equilibrium statistical mechanics,
\begin{equation}\label{2b}
  \calP(\bm{\eta};Y) \propto \exp \left[ -A(\bm{\eta};Y)/T  \right]\,  \delta\Bigg(\sum_{j=1}^N \eta_j-L\Bigg),
\end{equation}
\end{subequations}
where the temperature is measured in energy units. The equilibrium values of the extensions correspond to the maxima of $\calP(\bm{\eta};Y)$ or, equivalently, to the minima of the thermodynamic potential $A(\bm{\eta};Y)$ for the given length. We thus look for the minima of $A(\bm{\eta};Y)-FL=\sum_j [a(\eta_j;Y)-F\eta_j]$, where $F$ is a Lagrange multiplier with force dimensions that has to be calculated at the end of the procedure by imposing the length constraint, $F=F(L)$. From a physical point of view, $F$ is the force that must be applied in order to have the desired length. Thus, the relation $F=F(L)$ is the force-extension curve.   Because of the homogeneity and non-interaction assumptions for the modules, the equation determining the extension equilibrium value $\eta_j^\eq(T,F)$ is independent of the considered module,
\begin{equation}\label{3}
  -2 \alpha \eta^{(\kappa)} +4 \beta \left(\eta^{(\kappa)}\right)^3=\varphi, \quad \varphi\equiv F-F_c.
\end{equation}
In the metastability region given by,
\begin{equation}\label{3b}
  |\varphi|=|F-F_c|< \varphi_0=\left(\frac{2 \alpha}{3 \beta^{1/3}}\right)^{3/2},
\end{equation}
(\ref{3}) has three solutions $\eta^{(\kappa)}$, $\kappa=1,2,3$, with $\1<\2<\3$. They depend on the intensive variables $Y$ through $\{\alpha,\beta,F_c\}$, and also on $F$.  To keep our notation simple, we omit the dependence on $F$ and $Y$ from now on. The extensions $\1$ and $\3$ are locally stable because they correspond to local minima of $a_j-F\eta_j$, while $\2$ corresponds to a maximum and is therefore unstable. Then, each module can be either folded, with $\eta_j^\eq=\1<0$, or unfolded, with $\eta_j^\eq=\3>0$.  Both stable extensions $\1$ and $\3$ are increasing functions of $F$. Besides, the folded $\1$ (resp.\ unfolded $\3$) state also exists for $\varphi<-\varphi_0$ (resp.\ $\varphi>\varphi_0$). The equilibrium value of $\calA$ is
\begin{equation}\label{5}
  \calA^\eq(L;Y)=A(\bm{\eta}^\eq(L;Y);Y)=\!\sum_{j=1}^N a(\eta_j^\eq(L;Y);Y).
\end{equation}
Equation (\ref{3}) is equivalent to  $\left. \partial a/\partial \eta_j\right|_\eq=F$, $\forall j$, and thus it is straightforward to show that
\begin{equation}\label{6}
  \left(\frac{\partial\calA^\eq}{\partial L}\right)_Y=F.
\end{equation}

\begin{figure}[htbp]
\begin{center}
    \includegraphics[width=3.25in]{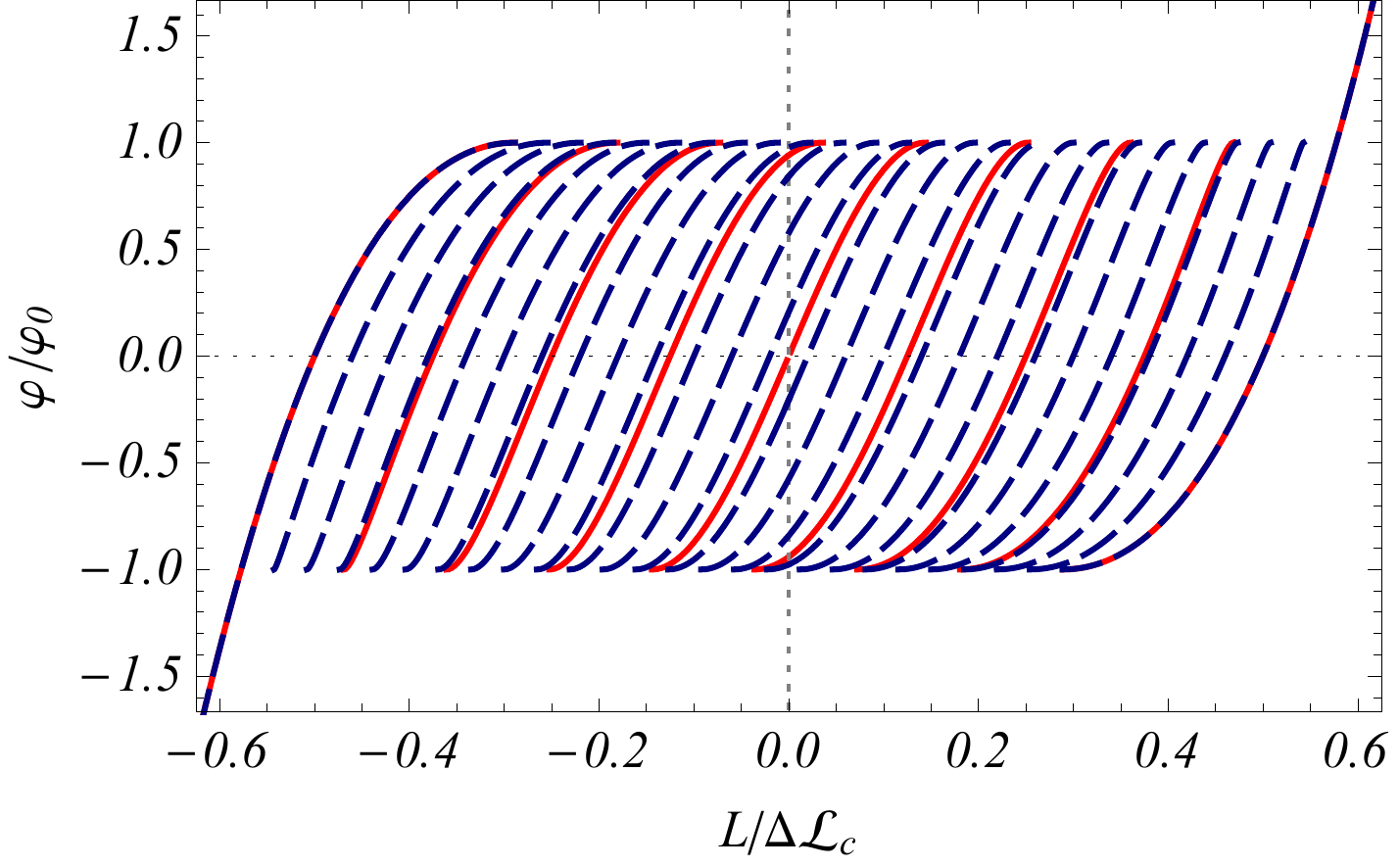}
    \caption{Normalized force-extension curves for
    (a) $N=8$ (solid red) and (b) $N=25$ (dashed blue). We plot the normalized force $\varphi/\varphi_0=(F-F_c)/\varphi_0$ vs.\ the normalized length $L/\Delta\calL_c$, where $\Delta \calL_c=\calL_N(F_c)-\calL_0(F_c)$ is the difference of lengths between the points at the first and last branches at the critical force $F_c$.
     Zero length corresponds to having half of the units unfolded at $F=F_c$. There are $N+1$ branches in the metastability region $|\varphi/\varphi_0|< 1$, with the number of unfolded units $J$ increasing from left to right.  The first ($J=0$) and last ($J=N$) branches are independent of $N$, because they correspond to all the units being folded and unfolded, respectively. Note that the branches become denser as $N$ increases.   }
    \label{ramas_eq}
\end{center}
\end{figure}

The $F-L$ curve has $N+1$ branches in the metastability region, see fig. \ref{ramas_eq}.  The $J$-th branch corresponds to $J$ unfolded modules and $N-J$ folded ones, $J=0,\ldots,N$, over which the equilibrium free energy and the $F-L$ curve are given by
\begin{equation}\label{4}
  \calA_J^\eq=(N-J)\auno +J \atres\!, \,
  \calL_J = (N-J) \1+J\3\!.
\end{equation}
We have defined $\auno\equiv a(\1)$, $\atres\equiv a(\3)$.  The equilibrium extensions $\1$ and $\3$ depend on $F$ and $Y$, so that $\calL_J$ represents the length as a function of $F$ and $Y$ over the $J$-branch. The constraint $L=\calL_J$ allows us to obtain its force extension curve $F_J=F_J(L,Y)$.

\begin{figure}[htbp]
\begin{center}
    \includegraphics[width=3.25in]{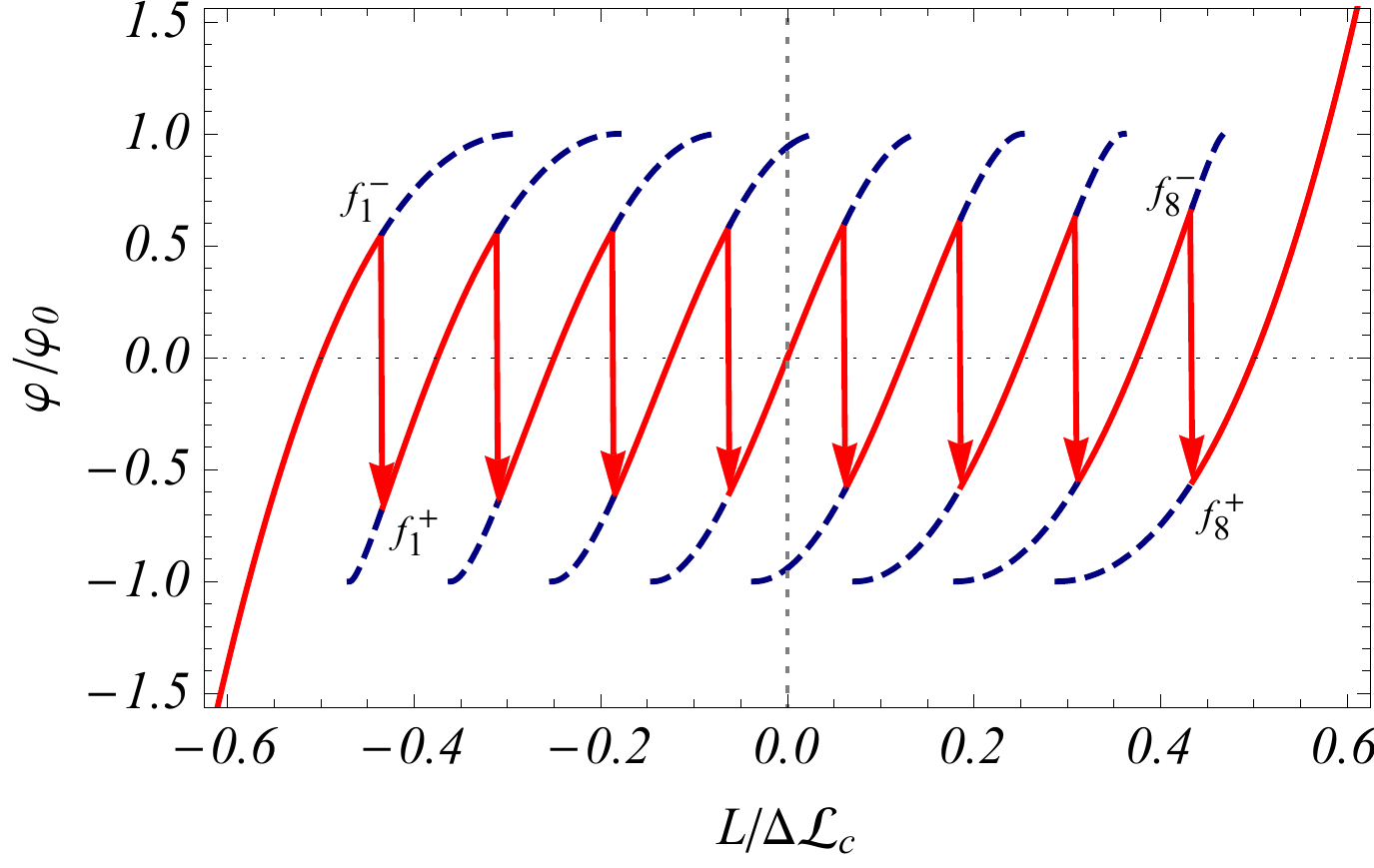}
    \caption{Equilibrium force rips in the $F-L$ curve for a system with $N=8$ domains. Different colors are used for the stable parts of the branches (solid red), metastable parts (dashed blue), and the force rips (red arrows).  If the length were increased in a quasi-static manner, the system would follow the solid red curve, with a series of first order transitions in the force (marked by the arrows). Note that the minimum values of these force rips $f_J^+$ increase with the number of unfolded units $J$, as observed in real AFM experiments with modular proteins, even though all the units are perfectly identical in the model. This is necessary in order to fulfill the continuity condition for the free energy.}
\label{force_rips}
\end{center}
\end{figure}

A natural question arises now: at a given length $L$, which is the most stable branch? The equilibrium probability of a given equilibrium configuration $\bm{\eta}^\eq$ is given by (\ref{2b}), so that
the difference of values of $\calA^\eq$ between branches governs the stability thereof. Therefore, the length $\ell_J$ at which there is a change in the relative stability of two consecutive branches, with $J-1$ and $J$ unfolded units, is determined by the equality of their respective free energies $\calA^\eq$. The corresponding forces $f_J^-\equiv F_{J-1}(\ell_J)$ and $f_J^+=F_J(\ell_J)$ over the branches with $J-1$ and $J$ unfolded units obey the system of two equations
\begin{equation}\label{7}
\left. \calA_{J-1}^\eq\right|_{f_J^-}\!=\left. \calA_{J}^\eq\right|_{f_J^+}, \quad
\left. \calL_{J-1}\right|_{f_J^-}\!=\left. \calL_{J}\right|_{f_J^+}.
\end{equation}
The force rips at $L=\ell_J$ are $N$ first-order equilibrium phase transitions because the thermodynamic potential $\calA^\eq$ is continuous at the transition, but $F=(\partial\calA^\eq/\partial L)_Y$ has a finite jump, from $f_J^-$ to $f_J^+<f_J^-$ at the $J$-th transition, as shown in Fig.\ \ref{force_rips}.
In fact, the following picture arises: there is a range of lengths in which the branches $J-1$ and $J$ coexist (see fig.\ \ref{ramas_eq}). In this coexistence region, eq.\ (\ref{6}) implies that
\begin{equation}\label{8}
  \left(\frac{\partial}{\partial L} \left[\calA^\eq_{J}-\calA^\eq_{J-1} \right]\right)_Y=F_J(L)-F_{J-1}(L)<0,
\end{equation}
because, at equal length values $L$, the force is larger on the branch with a smaller number of folded units, $F_J(L)<F_{J-1}(L)$, $\forall J$. Then the branch $J-1$ is the stable one and $J$ is metastable, $\calA^\eq_{J-1}<\calA^\eq_{J}$, for $L<\ell_J$. The situation reverses for $L>\ell_J$, and there cannot be more changes of stability because $\calA^\eq_{J}-\calA^\eq_{J-1}$ is monotonically decreasing as a function of $L$ in the metastability region (\ref{3b}). Each intermediate branch ($J=1,\ldots,N-1$) is thus stable between $\ell_J$ and $\ell_{J+1}$, that is, between $f_J^+$ and $f_{J+1}^-$ (see Fig. \ref{force_rips}). A  sawtooth pattern arises in the $F-L$ curve, with $N$ transitions between the $N+1$ branches at lengths $\ell_1,\ldots,\ell_{N}$.

\begin{figure}[htbp]
\begin{center}
\includegraphics[width=3.25in]{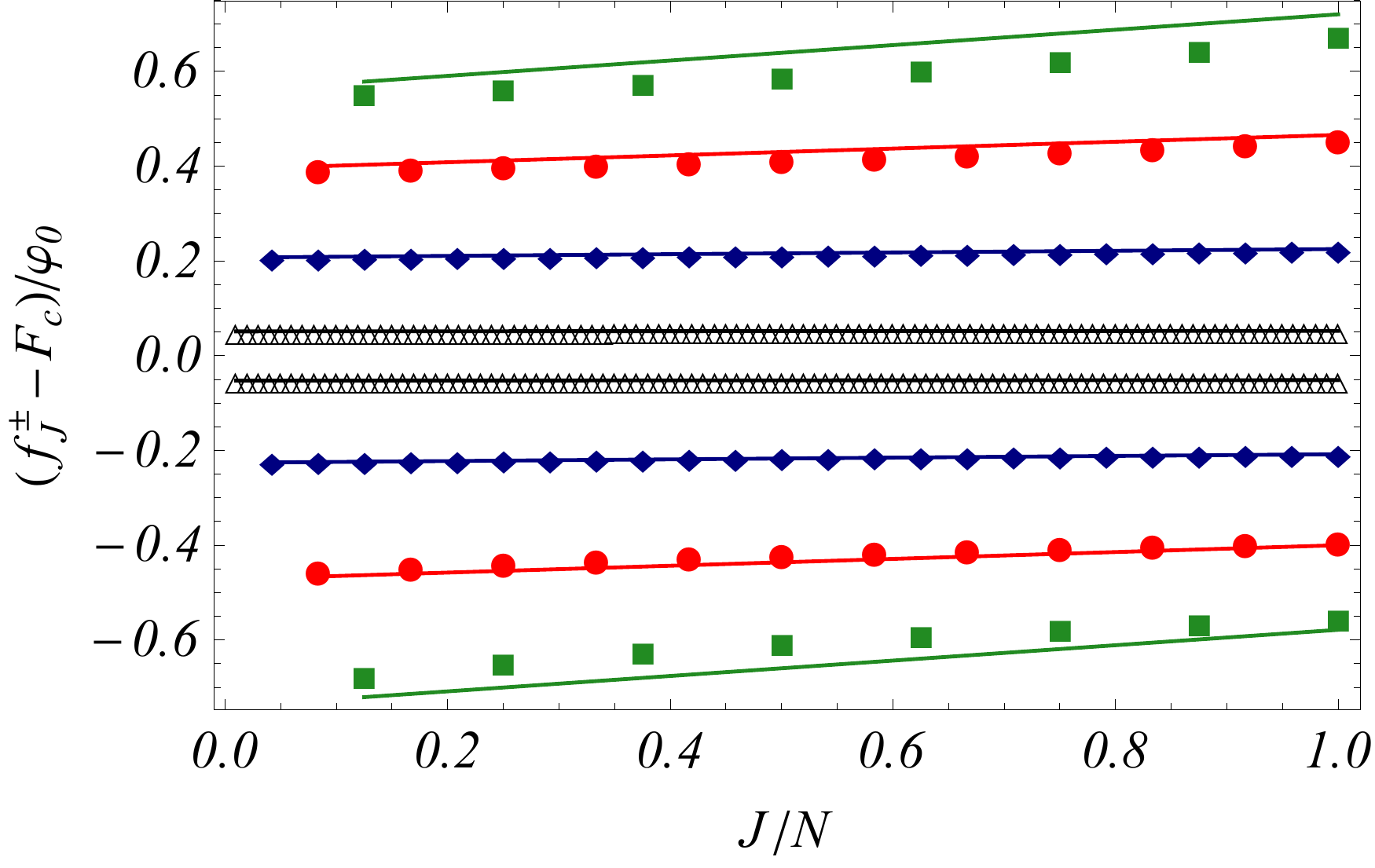}
\caption{Force rips as a function of the number of the fraction of unfolded units $J/N$. We compare the large $N$ theoretical prediction (\ref{9})
(lines) to the numerical solution of (\ref{7}) (symbols), for $N=8$ (squares), $N=12$ (circles), $N=24$ (diamonds), and $N=100$ (triangles). Note that the size of the force rips decreases with $N$. The theory correctly predicts the increases of $f_J^\pm$ with $J$ (for a fixed value of $N$), and the quantitative agreement is already remarkably good for $N=24$. }
\label{fig3v2}
\end{center}
\end{figure}

Finally, let us analyze the dependence of the force rips with the number of units $N$. For $N\gg 1$, the branches become dense (see fig. \ref{ramas_eq}) and the force rips sizes decrease accordingly. The free energy over each branch is proportional to $N$, while the difference of free energies (for a given value of the force) is of the order of unity; this means that the relative free energy change between consecutive branches is of the order of $N^{-1}$. Making use of eq.(8) and neglecting terms of order $N^{-3}$, we obtain
\begin{equation}\label{9}
 \frac{f_J^{\pm}-F_c\!}{\varphi_0}=\mp\frac{3\sqrt{3}}{N} \!\left(1\mp\frac{r_J}{N}\right)\!, \;\; r_J\!=\left.\frac{\calL_{J-1}+\calL_{J}}{\calL_N-\calL_0}\right|_{F_c}\! .
\end{equation}
We have that $r_J$ increases linearly with $J$, see (\ref{4}), and thus so do both $f_J^-$ and $f_J^+$. For very large $N$, the term proportional to $r_J$ is negligible and thus the force rips become independent of $J$ and symmetrical with respect to $F_c$, $f_J^\pm-F_c\sim \mp 3\sqrt{3}\varphi_0/N$. This is consistent with the behavior observed in nucleic acids \cite{rit06jpcm,HBFSByR10}, whose number of units is much larger than that typical of modular proteins. In Fig. \ref{fig3v2} we plot both the height of the force rips obtained by numerically solving (\ref{7}) and the theoretical prediction (\ref{9}). Note that $\lim_{N\to\infty}f_J^\pm=F_c$: force fluctuations disappear and all the units of the system unfold simultaneously at the critical force $F_c$. This would be the  expected behavior had we controlled the force instead of the length \cite{BCyP13}.

We have seen that under a simple but quite general model the fingerprints of $F-L$ curves in biomolecules can be understood within a purely equilibrium statistical mechanics description. We stress that most of these key features are already present for perfectly homogeneous systems at equilibrium in the absence of thermal fluctuations when the monomers' free energy has two stable extensions (folded/unfolded). This causes the system to have multiple free energy branches in a certain force range. In fact, the fingerprints of the $F-L$ curve  stem from (i) branch multistability  (ii) the continuity of the relevant thermodynamic potential in consecutive branches. The latter implies, under length controlled conditions, abrupt changes of force (force rips) at certain lengths at which consecutive branches change their mutual stability.

For systems with a small number of units, like modular proteins, the forces at the rips increase with the number of unfolding events, even though all the units are identical.  This is in contrast with the usual explanation of this behavior in the literature, where it has been usually attributed to slight differences between the modules of the polyprotein \cite{MyD12,FOCMyF99,COFMBCyF99,RFyG98}. Of course, in real experiments all the modules are not perfectly identical, but the force increases at the rips are definitely not a signature of their heterogeneity. On the other hand, the amplitude of these force rips decreases as the number of units in the system increases, explaining why the rips are observed in a cleaner way for modular proteins than for nucleic acids. In the latter, the number of \textit{units} is much larger and the smaller rips are thus much more prone to be affected by perturbations present in real experiments, such as thermal fluctuations (noise), heterogeneity of the units, non-quasi-static increase of the length, etc.

Of course, the laboratory conditions of SMEs are not always such that the biomolecule is at thermodynamic equilibrium. It is worth  noting that hysteretic behavior is to be expected when the molecule is unfolded and afterwards refolded at a finite rate.  In a pulling (resp.\ pushing) process, the dynamical $F-L$ curve is moved to higher (resp.\ lower) forces with respect to the quasi-static limit, because the system sweeps different regions of the metastable branches \cite{PCyB12,Ka12,BCyP13}. However, the separation from the quasi-static behavior will be small if the characteristic time of the pulling process is large as compared with the time that the system needs to surpass the free energy barrier at the critical force $F_c$. Furthermore, even at equilibrium, thermal fluctuations, short-range interactions between modules and the unavoidable heterogeneity in real experiments will modify the results presented here. Finally, the simplicity of the proposed double-well free energy for each monomer prevents us from doing quantitative comparisons to experimental results. For that, more realistic free energies should be used, like the one arising in the worm-like chain (WLC) model that describes polymer elasticity \cite{BMSyS94,MyS95}.

\acknowledgments
This work has been supported by the Spanish Ministerio de Econom\'\i a y Competitividad grants FIS2011-24460 (AP), FIS2011-28838-C02-01 (LLB), and  FIS2011-28838-C02-02 (AC). AP thanks J. Javier Brey for really helpful discussions.

\end{document}